

	\footline={\ifnum\pageno=1\firstfootline\else\otherfootline\fi}
\def\firstfootline{\rm\hss\folio\hss}
\def\otherfootline{\hfil}

\font\twelvebf=cmbx10 scaled\magstep 1
 1
 1

\font\tenrm=cmr10
\font\tenit=cmti10

\parindent=1.2pc
\magnification=\magstep1
\hsize=6.0truein
\vsize=8.6truein
\nopagenumbers

\input jytex.tex
%
%


\def\h{\hbox to .5cm{\hfill}}
\def\hof{\hbox to .15cm{\hfill}}
\def\hi{\hbox to .2cm{\hfill}}
\def\htwo{\hbox to .2cm{\hfill}}
\def\htf{\hbox to .35cm{\hfill}}

\def\ha#1{\n\hbox to .6cm{\n {#1}\hfill}}
\def\hb#1{\indent\hbox to .7cm{\n {#1}\hfill}}
\def\hc#1{\indent\hbox to .7cm{\hfill}\hbox to 1.1cm{\n {#1}\hfill}}
\def\hd#1{\indent\hbox to 1.8cm{\hfill}\hbox to 1.4cm{\n {#1}\hfill}}

\def\hj{{\hbox to 0.06truecm{\hfill}}}


\def\pr#1{[\putref{#1}]}

\def\n{\noindent}
\def\no{\noindent}
\def\ref{\reference}

\def\undbib{\underbar {\hbox to 2cm{\hfill}}, }
%
%

\def\nall{ $N= 4$, 2, 1, 0 ST-SUSY }
\def\nall1{  $N= 4$, 2,    0 ST-SUSY }

%
%

\def\bS{ {\bmit S}}


%
%

\def\half{{1\over 2}}

\def\third{{1\over 3}}

\def\twothird{{2\over 3}}

\def\fourth{{1\over 4}}

\def\threefourth{{3\over 4}}

\def\fifth{{1\over 5}}

\def\twofifth{{2\over 5}}
\def\threefifth{{3\over 5}}
\def\fourfifth{{4\over 5}}

%
%

%
%

\begin{ignore}

\font\twelveBbb=msym10 scaled \magstep1
\font\nineBbb=msym9
\font\sevenBbb=msym7
\newfam\Bbbfam
\textfont\Bbbfam=\twelveBbb
\scriptfont\Bbbfam=\nineBbb
\scriptscriptfont\Bbbfam=\sevenBbb

 \def\Z{{\Bbb Z}}
\end{ignore}
\def\Z{Z\!\!\!Z}

\def\va{\vskip .2truecm}

\topmargin= 1truein\vsize=8.6truein
\leftmargin=1truein\hsize=6.0truein
\baselinestretch=1000

\foot={\hfil\normalfonts\numstyle\pagenum\hfil}
\head={\hfil}

\footnotenumstyle{symbols}
\footnotenum= 0
\sectionnum=0\equationnum=0
\sectionnumstyle{arabic}
\equationnumstyle{arabic}

\pagenumstyle{arabic}

{\rightline{\hbox to 4.5cm{\vtop{\hsize= 4.5cm
\baselinestretch=960\footnotefonts
\hfill\\
OHSTPY-HEP-T-95-003\\
DOE/ER/01545-643\\
hep-th/9506006\\
June 1995}}{\hbox to .35cm{\hfill}}}}

\vskip 1.0 truecm

\centerline{\twelvebf WHAT'S NEW IN STRINGY SO(10) SUSY-GUTS}
\baselineskip=15pt

\centerline{\tenrm GERALD B.~CLEAVER}
\baselineskip=13pt
\centerline{\tenit Dept. of Physics,
The Ohio State University}
\baselineskip=13pt
\centerline{\tenit Columbus, Ohio 43210-1106}
\centerline{\tenrm E-mail: gcleaver@ohstpy.bitnet}
\vglue0.4cm

\baselineskip=12pt

As a stepping-stone in our search for string-derived three-generation SO(10)
SUSY-GUTs, we investigated
the six distinct gravitino generators $\bS_i$
(see Table 1)
in heterotic free fermionic strings and applied
all consistent combinations of unique GSO
projections (GSOPs) to them.$^{1}$
For each gravitino generator,
we determined how many of the initial $N=4$ spacetime
supersymmetries (ST-SUSYs)
can survive various combinations of GSOPs.
Our findings can be summarized as follows (noting that a $\Z_n$ twisted
boundary vector (BV) contains components of the form ${2a\over n}$ where
$a$ and $n$ are relative primes in at least one  component):
\item{1.} Only left-moving (LM) $\Z_2$, $\Z_4$, and $\Z_8$ twists
that correspond to automorphisms of SU(2)$^6$
are consistent with $N=1$ in free fermionic models.
All other LM $\Z_n$ twists obviate $N=1$.
Thus, neither gravitino generators $\bS_5$ and $\bS_7$
(both with $\Z_6$ twists), nor $\bS_{10}$ (with $\Z_{10}$ twists)
can produce $N=1$ ST-SUSY.
$\bS_5$ and $\bS_7$ only result in $N=4$, 2, or 0, whereas
$\bS_{10}$ yields $N=4$ or 0.
\item{2.} $N=1$ ST-SUSY is possible for $\bS_1$, $\bS_3$, and $\bS_9$.
Six general categories of GSOP sets
for $\bS_1$, three for $\bS_3$, and one for $\bS_9$
lead, respectively, to $N=1$. The GSOPs in these sets originate
from LM BVs with $\Z_2$, $\Z_4$, and $\Z_8$ twists.

We have completely
classified the ways by which the number of ST-SUSYs
in heterotic free fermionic strings
may be reduced from $N=4$ to the phenomenologically preferred $N=1$.
This means that the set of LM BVs
in any free fermionic model  with $N=1$ ST-SUSY
must be reproducible from one of the three specific
gravitino sectors, $\bS_1$, $\bS_3$, or $\bS_9$, combined with one of our
left-moving BV sets whose GSOPs reduce the initial $N=4$ to $N=1$.
The only variation from our BVs that true $N=1$ models could have
(besides trivial reordering of BV components) is
some component sign changes, which we have shown
do not lead to new, physically distinct models.

To this date, only the gravitino generator $\bS_1$
has been used in actual $N=1$ models.
Reduction to $N=1$ ST-SUSY has been accomplished through
GSOPs from the NAHE set of boundary vectors.$^2$
Thus, our new results should be especially useful for
model building
when the NAHE set may be inconsistent with other properties
specifically desired in a model.
This appears to be the situation with regard to
current searches for consistent three generation
SO(10) level-2 models. Initial results of this search
were discussed in refs.~3 and 4.
Attempts to simultaneously produce
$N=1$ ST-SUSY and a three-generation SO(10) level-2 grand
unified theory,
using for left-movers $\bS_1$, the NAHE set, and one or two
additional BVs containing some non-integer components,
were initially thought to be successful.
However, it was later discovered that
the extra non-integer BVs required did not correspond to proper
SU(2)$^6$ automorphisms and, therefore, resulted in
additional sectors containing tachyonic spacetime fermions.

Although we have found nine new solutions for generating $N=1$
ST-SUSY, it remains to be shown that these are all
physically unique from the standard NAHE set of GSOPs and boundary vectors.
That is, we must check for instances when an $N=1$ model that does
not use the standard NAHE solution is phenomenologically
equivalent to an $N=1$ model that does.
Identities relating partition functions for
products of (anti)periodic worldsheet
fermions to those for certain products of
complex fermions have been derived.$^{5,6}$
These identities will be used to test for possible physical equivalences.
Following this, we will investigate which of our physically
unique LM $N=1$ ST-SUSY solutions may be consistent with
three-generation SO(10) level-2 GUT models.

\vglue .4truecm
\no {\bf TABLE 1.}
\vglue .3truecm
\no
\hbox to 2.4truecm{\bf Class\hfill}
\hskip .2truecm
\hbox to 12.5truecm{\hfill \bf Unique Massless Gravitino Boundary Vectors
\hfill}

\no
\underline{\hbox to 2.4truecm{\hfill $\phantom{*}$ \hfill}}
\hskip .2truecm
\underline{\hbox to 12.5truecm{\hfill $\phantom{*}$\hfill}}

\no
\hbox to 2.4truecm{$1\cdot1\cdot1\cdot1\cdot1\cdot1$\hfill}
\hskip .2truecm
\hbox to 1.1truecm{\hfill $\bS_1$ \hfill $=$}
\hskip .2truecm
\hbox to 11.1truecm{$\{ 1,1\quad (1; 0, 0)^6 \}$ \hfill}\hfill

\va
\no
\hbox to 2.4truecm{$2\cdot2\cdot1\cdot1$\hfill}
\hskip .2truecm
\hbox to 1.1truecm{\hfill $\bS_3$ \hfill $=$}
\hskip .2truecm
\hbox to 11.1truecm{$\{ 1,1\quad (0,1; -\hat{\half}, \hat{\half})^2\quad
                                      (1; 0, 0)^2 \}$ \hfill}\hfill

\va
\no
\hbox to 2.4truecm{$3\cdot1\cdot1\cdot1$\hfill}
\hskip .2truecm
\hbox to 1.1truecm{\hfill $\bS_5$ \hfill $=$}
\hskip .2truecm
\hbox to 11.1truecm{$\{ 1,1\quad
                 (\hat{\third},1; -\hat{\twothird}, 0,0,\hat{\twothird})\quad
                                      (1; 0, 0)^3 \}$ \hfill}\hfill

\va
\no
\hbox to 2.4truecm{$3\cdot3$\hfill}
\hskip .2truecm
\hbox to 1.1truecm{\hfill $\bS_7$ \hfill $=$}
\hskip .2truecm
\hbox to 11.1truecm{$\{ 1,1\quad
                 (\hat{\third},1; -\hat{\twothird}, 0,0,\hat{\twothird})^2
                                                              \}$ \hfill}\hfill

\va
\no
\hbox to 2.4truecm{$4\cdot2$\hfill}
\hskip .2truecm
\hbox to 1.1truecm{\hfill $\bS_9$ \hfill $=$}
\hskip .2truecm
\hbox to 11.1truecm{$\{ 1,1\quad
   (0,\hat\half, 1; -\hat{\threefourth}, -\hat{\fourth},
                     \hat{\fourth},  \hat{\threefourth})\quad
                                      (0,1; -\hat{\half}, \hat{\half})
                                                              \}$ \hfill}\hfill

\va
\no
\hbox to 2.4truecm{$5\cdot1$\hfill}
\hskip .2truecm
\hbox to 1.1truecm{\hfill $\bS_{10}$ \hfill $=$}
\hskip .2truecm
\hbox to 11.1truecm{$\{ 1,1\quad
   (\hat{\fifth},\hat{\threefifth},1; -\hat{\fourfifth},-\hat{\twofifth},0,0,
                                      \hat{\twofifth},  \hat{\fourfifth})\quad
                                      (1; 0, 0)\}$ \hfill}\hfill

\vglue .4truecm
\no{\bf Acknowledgements}
\vglue .2truecm

Our sincere
thanks to Henry Tye, Jorge Lopez, and Kajia Yuan for helpful discussions.

\vglue .2truecm
\no{\bf References}

\def\np#1{  {\it Nucl.~Phys.~}{\bf B#1}  }
\def\pl#1{  {\it Phys.~Lett.~}{\bf B#1}  }
\def\pr#1{  {\it Phys.~Rev.~}{\bf D#1}   }

\item{1.}{G.~Cleaver, {\it ``Number of Supersymmetries in Free
                 Fermionic Strings,''}
                 OHSTPY-HEP-T-95-004; DOE/ER/01545-643;
                 hep-th/9505080.}
\item{2.}{A.~Faraggi, \np{387} (1992) 239;\\
A.~Faraggi and D.V.~Nanopoulos, \pr{48} (1993) 3288.}
\item{3.}{G.~Cleaver, {\it ``GUT's With Adjoint Higgs Fields
                 From Superstrings,''} OHSTPY-HEP-T-94-007. To appear in the
                 Proceedings of PASCOS '94, Syracuse, N.Y.}
\item{4.}{S.~Chaudhuri, S.~Chung, and J.~Lykken,
     {\it ``Fermion Masses from Superstring Models with Adjoint Scalars,''}
     Fermilab-PUB-94/137-T; NSF-ITP-94-50.}
\item{5.}{D.~Bailin, D.~Dunbar, and A.~Love, \np{330} (1990) 124;\\
     D.~Bailin and A.~Love, \pl{292} (1992) 315.}
\item{6.}{G.~Cleaver, Unpublished research.}
\newpage

\bye